\documentclass[onecolumn,A4]{article}
\usepackage{cite}
\usepackage{anysize}
\usepackage{mathrsfs}
\usepackage{amsmath}
\usepackage{amssymb}
\usepackage{graphicx}
\usepackage{subfigure}
\usepackage{amssymb}
\usepackage{amsthm}
\pdfoutput=1
\usepackage{pgf}
\usepackage{pdfpages}
\DeclareGraphicsExtensions{.jpeg}

\begin{document}

\vskip 1cm
\marginsize{3cm}{3cm}{3cm}{1cm}

\begin{center}
{\bf{\huge 3D Simulation of Electron and Ion Transmission of GEM-based Detectors}}\\
~\\
Purba Bhattacharya$^{a*1}$, Bedangadas Mohanty$^a$, Supratik Mukhopadhyay$^b$, Nayana Majumdar$^b$, Hugo Natal da Luz$^c$\\
{\em $^a$ School of Physical Sciences, National Institute of Science Education and Research, Jatni, Bhubaneswar, India}\\
{\em $^b$ Applied Nuclear Physics Division, Saha Institute of Nuclear Physics, Kolkata, India}\\
{\em $^c$ High Energy Physics and Instrumentation Center, Instituto de F$\acute{i}$sica, Universidade de S$\tilde{\mathrm{a}}$o Paulo, Brazil}\\
~\\
~\\
~\\
~\\
~\\
{\bf{\large Abstract}}
\end{center}

\noindent Time Projection Chamber (TPC) has been chosen as the main tracking
system in several high-flux and high repetition rate experiments.
These include on-going experiments such as ALICE and
future experiments such as PANDA at FAIR and ILC.
Different $\mathrm{R}\&\mathrm{D}$ activities were carried out
on the adoption of Gas Electron Multiplier
(GEM) as the gas amplification stage of the ALICE-TPC upgrade
version.
The requirement of low ion feedback has been established through
these activities.
Low ion feedback minimizes distortions due to space charge
and maintains the necessary values of detector gain and energy resolution.
In the present work, Garfield simulation framework has been
used to study the related physical processes
occurring within single, triple and quadruple GEM detectors.
Ion backflow and electron transmission of quadruple GEMs,
made up of foils with different hole pitch under different
electromagnetic field configurations (the projected
solutions for the ALICE TPC) have been studied.
Finally a new triple GEM detector
configuration with low ion backflow fraction and good electron
transmission properties has been proposed as a simpler GEM-based
 alternative suitable for TPCs for future collider experiments.

\vskip 1.5cm
\begin{flushleft}
\end{flushleft}

\vskip 1.5in
\noindent
{\bf ~$^*$Corresponding Author}: Purba Bhattacharya\\
\noindent {\bf ~$^1$} Presently at Department of Particle Physics and AstroPhysics, Weizmann Institute of Science, Herzl St. 234, Rehovot - 7610001, Israel\\

E-mail: purba.bhattacharya85@gmail.com

\newpage

\section{Introduction}

The physics processes aimed at various on-going and future high energy 
and particle physics experiments, have pushed the detector requirements 
to an unprecedented level.
Owing to the enormous particle multiplicity per event, 
these requirements include good momentum resolution, high 
jet energy resolution, excellent particle identification
and ability to cope with the harsh radiation environments.
Time Projection Chambers (TPC) \cite{TPC}, due to their low material 
budget and excellent pattern recognition capabilities, 
are often used for three-dimensional tracking and 
identification of charged particles.
They constitute the main tracking system in many on-going experiments, 
such as ALICE \cite{ALICE} and are proposed to be used for 
several future experiments such as PANDA \cite{PANDA} and ILC \cite{ILC}. 
Since the ALICE experiment is an on-going one planning for a significant 
upgrade within a few years  time scale, extensive $\mathrm{R}\&\mathrm{D}$ 
has been carried out for the upgrade part of its TPC.

ALICE (A Large Ion Collider Experiment) is one of the 
general-purpose heavy-ion experiments at the Large Hadron Collider 
(LHC) which is designed to study the physics of strongly interacting 
matter and the Quark Gluon Plasma (QGP) in nucleus-nucleus collisions. 
In order to identify all the particles that are coming out of the 
QGP, ALICE is using a set of 18 detectors that gives 
information about the mass, the velocity and the electrical sign 
of the particles. 
A significant increase of the LHC luminosity for heavy ions is expected 
in RUN 3 after Long Shutdown 2 (LS2), leading to collision rates of 
about 50 kHz for Pb-Pb collisions. 
This implies a substantial enhancement of the sensitivity to a number 
of rare probes that are key observable for the characterization of 
strongly interacting matter at high temperature. 
A continuous ungated mode of operation is the only way to run the 
TPC in 50 kHz Pb-Pb collisions.

The time necessary to evacuate the ion charge (created 
in the amplification process) from the detector volume is relatively high
for the current Multi Wire Proportional Chamber (MWPC) based readout
of the present ALICE-TPC. 
These ions drift back into the TPC volume, create local perturbations 
in the electric field and, thus, affect the drift behavior of the 
electrons from a later track.
This ion feedback problem restricts the use of MWPCs in high rate
experiments. 
Although this problem can be solved by using an additional plane of gating
grid, it leads to an intrinsic dead time for the TPC, implying 
a rate limitation of the present TPC.

To fully exploit the scientific potential of the LHC at high-rate 
Pb-Pb collisions, the ALICE collaboration plans an upgrade 
of many sub-detectors, including the central tracker 
\cite{ALICE-Upgrade, ALICE1}. 
Different $\mathrm{R}\&\mathrm{D}$ activities have been carried out 
and converged to the adoption of Gas Electron Multiplier 
(GEM) \cite{GEM} as the gas amplification stage of the ALICE-TPC upgrade 
version \cite{ALICE2} while retaining the present tracking and particle
identification capabilities of the TPC via measurement of the 
specific energy loss (dE/dx).
The new readout chambers will employ stacks of four GEM foils
for gas amplification and anode pad readout. 
The configuration consists of a
combination of standard (S) and large hole pitch (LP) GEM foils, i.e.,
S-LP-LP-S.
Such quadruple GEM stacks have been found to provide sufficient ion blocking 
capabilities at the required gas gain of $2000$ in 
$\mathrm{Ne}/\mathrm{CO_2}/\mathrm{N_2}$ ($90/10/5$). 
However, further optimization of the experimental parameters (geometry,
electrostatic configuration, gas composition, material used to build the
detector components) can minimize distortion due to space charge by
reducing ion feedback in the drift volume \cite{IBF} and larger 
signals through improved electron transmission.

In this work, we have tried to develop a thorough understanding of GEM-based
detectors from this point of view and made attempts to explore the
appropriateness / suitability of these detectors in the context of the TPC
in general.
Extensive numerical simulations have been carried out to
estimate the effects of detector geometry, electric field configurations and
magnetic field on electron transmission and ion backflow fraction.
To begin with, single GEM configurations have been studied in detail 
and compared with available experimental data.
A good understanding of this device 
has allowed us to deal with the quadruple GEM configuration with 
relative ease.
The numerical results for the quadruple GEM have been also compared with 
the available experimental data of ALICE TPC.
Finally, we have worked on a new configuration of a triple GEM 
detector which allows low ion backflow fraction despite providing 
good electron transmission and may be suitable for the TPCs in 
future collider experiments.
The stability of the detector behavior and the discharge probability 
are very important for the operation and most importantly they are 
affected by the geometry and field configurations. 
In the present simulation, all these issues are not taken into account.
Thus, the proposed solutions may need to be evaluated as 
regard to the overall stability of the detector.

\section{Simulation Tools}

The Garfield \cite{Garfield1, Garfield2} simulation framework has 
been used in the present work.
The 3D electrostatic field simulation has been carried out 
using neBEM (nearly exact Boundary Element Method) 
\cite{neBEM1, neBEM2, neBEM3} toolkit.
Besides neBEM, HEED \cite{HEED1, HEED2} has been used for primary
ionization calculation and Magboltz \cite{Magboltz1, Magboltz2} for 
computing drift, diffusion, Townsend and attachment coefficients.

\section{Simulation Models}

\begin{table}[hbt]
\caption{Design parameters of GEM-based detectors.}\label{GEMdesign}
\begin{center}
\begin{tabular}{|c|c|}
\hline
Polymer substrate & $50~\mu\mathrm{m}$ \\
\hline
Copper coating thickness & $5~\mu\mathrm{m}$ \\
\hline
Hole diameter (copper layer) & $70~\mu\mathrm{m}$ \\
\hline
Hole diameter (Polymer substrate) & $50~\mu\mathrm{m}$ \\
\hline
Hole to hole pitch & $140~/~280~\mu\mathrm{m}$ \\
\hline
Drift Gap & $3~\mathrm{mm}$ \\
\hline
Transfer gap 1 & $2~\mathrm{mm}$ \\
\hline
Transfer gap 2 & $2~\mathrm{mm}$  \\
\hline
Transfer gap 3 & $2~\mathrm{mm}$  \\
\hline
Induction gap & $2~\mathrm{mm}$ \\
\hline
\end{tabular}
\end{center}
\end{table}

\begin{figure}[hbt]
\centering
\subfigure[]
{\label{Area-1}\includegraphics[scale=0.25]{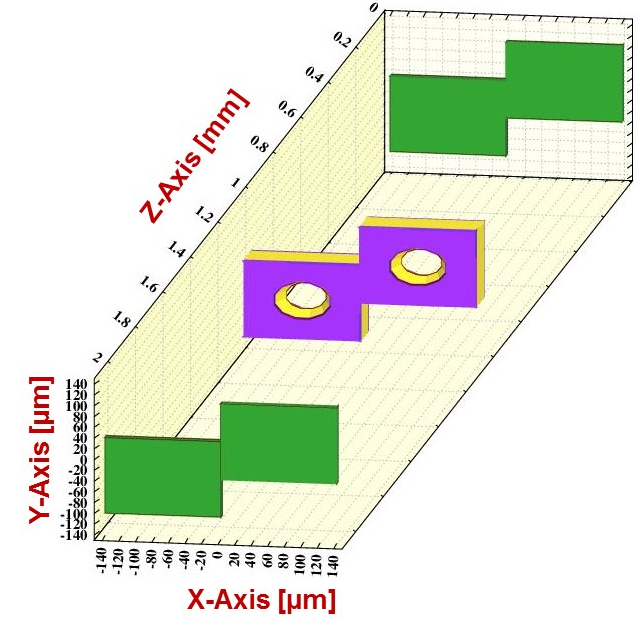}}
\subfigure[]
{\label{Area-2}\includegraphics[scale=0.25]{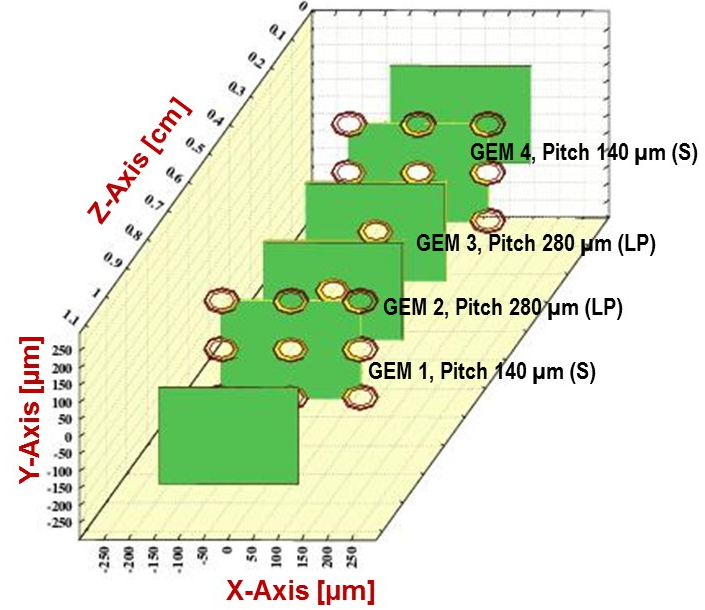}}
\subfigure[]
{\label{Area-3}\includegraphics[scale=0.25]{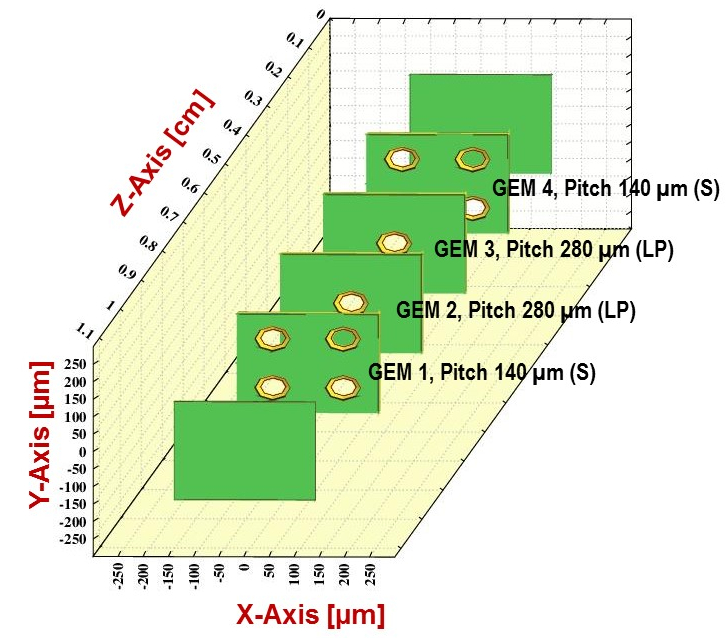}}
\caption{Model for (a) single and quadruple GEM with (b) aligned and (c) misaligned holes.}
\label{Area}
\end{figure}

The design parameters of GEM-based detectors considered in the numerical 
work, are listed in Table~\ref{GEMdesign}.
The model of a basic GEM cell, built using Garfield, is shown in Fig.~\ref{Area-1}.
It represents a GEM foil, having two bi-conical shaped holes placed in a
staggered manner along with a readout anode and a drift plane on either 
sides of the foil.
The distance between top surface of the GEM and the drift plane is called 
the drift gap whereas that between the lower surface and the readout 
plate is named induction gap.
The GEM foil separates these two volumes and is responsible for the 
transfer and amplification of the primary electrons generated in the 
drift volume. 
A potential difference $\mathrm{V_{Drift}}$ and $\mathrm{V_{Induction}}$ 
are maintained in the drift volume and the induction volume, 
respectively. 
The electric fields, both in the drift ($\mathrm{E_{Drift}}$) and 
induction ($\mathrm{E_{Induction}}$) volumes, are uniform and the
magnitudes have been kept at a value to meet the requirements of 
the electron drift and diffusion only. 
The large potential difference ($\mathrm{V_{GEM}}$) between 
the upper and lower GEM electrodes creates a strong field 
inside the holes ($\mathrm{E_{GEM}}$) which causes the 
amplification of the primary electrons.

In comparison to single GEM, in case of multi GEM detector, several GEM foils
are placed in between the drift and the read-out plane.
The naming scheme used in this work numbers the foils in the order of the
passage of electrons coming from the drift region.
The first GEM after the drift plane is called GEM 1 and the others 
are GEM 2, GEM 3 and so on.
The gap in between GEM 1 and 2 is called Transfer gap 1 and that between GEM
2 and 3 is called Transfer gap 2 etc.
The field in the transfer gap is uniform and the magnitudes
have been kept in a range suitable for the requirements of electron 
drift and diffusion.
For example, the simulation models of two different quadruple GEM devices are
shown in Fig.~\ref{Area}.
Among the four foils, GEM 1 and GEM 4 have the pitch of $140~\mu\mathrm{m}$
(denoted as S), whereas the middle two foils have a larger pitch of
$280~\mu\mathrm{m}$ (denoted as LP).
This arrangement is denoted as S-LP-LP-S.
In the first case (QGemI), the central hole of the basic unit from
all the four GEM foils are perfectly aligned (Fig.~\ref{Area-2}).
In the other case (QGemII), as shown in Fig.~\ref{Area-3}, the
first and the last foils (S) are aligned with each other whereas
the second and third foils (LP) are misaligned with them.
The basic cell structure then has been repeated along both positive
and negative X and Y-Axes to represent a real detector.
With the help of these models, the field configuration of the detectors
have been simulated using appropriate voltage settings.
These are followed by the simulation of electron transmission and
ion backflow fraction in $\mathrm{Ne}/\mathrm{CO_2}/\mathrm{N_2}$
(90/10/5) gas mixture.

For estimating electron transmission within a GEM detector, 
electron tracks generated by $5.9~\mathrm{keV}$ photon have been 
considered in the drift volume.
The primary electrons created in the drift region are then 
made to drift towards the GEM foil using the Microscopic tracking 
routine \cite{Garfield1}.
In this procedure, a typical drift path proceeds through millions
of collisions and each collision can be classified as elastic or
inelastic collision, excitation, ionization, attachment etc.

The electrons during their drift produce avalanche inside the GEM foil.
For this calculation Monte Carlo routine has been used.
The procedure first drifts an initial electron from the specified 
starting point. 
At each step, a number of secondary electrons is produced according 
to the local Townsend and attachment coefficients and the newly 
produced electrons are traced like the initial electrons.
In parallel, the ion drift lines are also traced. 
The primary ions in the drift region and the ions created in the avalanche
have been considered for the estimation of the backflow fraction.

\section{Results}

\subsection{Electron Transmission}

Electron transmission can be presented as a function of two mechanisms:
electron focusing and transverse diffusion.
The field configuration has a strong impact on electron focusing.
Due to the high field gradient between the drift volume and the GEM hole, 
the field lines are compressed, resulting in a characteristic funnel 
shape. 
The decrease of $\mathrm{E_{GEM}}$ for a particular $\mathrm{E_{Drift}}$ 
or the increase of $\mathrm{E_{Drift}}$ at a fixed $\mathrm{E_{GEM}}$ 
affects the funneling, resulting in the termination of the field 
line on the top surface of the GEM foil.
Again, the ratio between the $\mathrm{E_{GEM}}$ and $\mathrm{E_{Induction}}$ 
controls the field lines inside the GEM foil as well as in the 
induction volume.
Since $\mathrm{E_{Induction}}$ is lower than the field inside the GEM hole, the 
field lines emerging from the hole spread uniformly and finally end at 
the readout plane. 
Depending on the field ratio, the field lines emerging from the holes 
spread further away, promoting an increase of the number of electrons, 
while some field lines end on the bottom copper surface of the GEM foil.
In order to ensure the collection of a good percentage of
electrons on the readout plate, a proper optimization of the field in 
these three different regions is necessary.
Other important parameters such as attachment, diffusion depend on 
the gas mixture composition and $\mathrm{E/p}$. 
All these factors have important role in determining the final 
transmission.

\begin{figure}[hbt]
\centering
\includegraphics[scale=0.32]{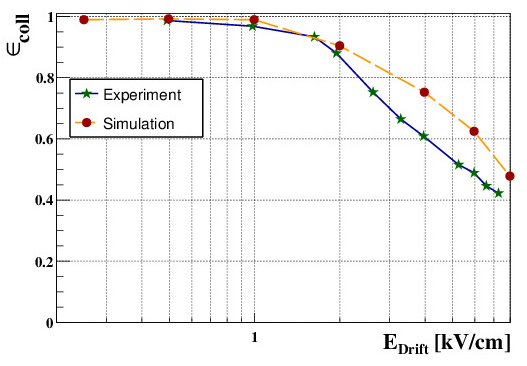}
\caption{The variation of $\epsilon_{{coll}}$ with ${\mathrm{E_{Drift}}}$. A comparison between the experimental data from \cite{Bachmann} and simulation results is shown here. The experimental details are described in the text.}
\label{Transparency-Comparison}
\end{figure}

\subsubsection{Single GEM}
\label{sec:trans}

For a single GEM detector, the total electron transmission ($\epsilon_{{tot}}$) 
can be identified as the multiplication of two efficiencies, the 
collection efficiency 
($\epsilon_{{coll}}$) and the extraction efficiency ($\epsilon_{{ext}}$).
The collection efficiency has been defined as:

\begin{eqnarray}
\epsilon_{{coll}}~=~{{\mathrm{\#~electrons~that~reach~inside~the~GEM~foil}} \over {\mathrm{\#~electrons~created~in~drift~volume}}}
\end{eqnarray}

\noindent The extraction efficiency has been defined as:

\begin{eqnarray}
\epsilon_{{ext}}~=~{{\mathrm{\#~electrons~that~reach~the~readout~plane}} \over {\mathrm{\#~electrons~that~reach~inside~the~GEM~foil}}}
\end{eqnarray}

Finally, the total transmission can be defined as:

\begin{eqnarray}
\epsilon_{{tot}}~=~{{\mathrm{\#~electrons~that~reach~the~readout~plane}} \over {\mathrm{\#~electrons~created~in~drift~volume}}}
\end{eqnarray}

We compared our numerical estimates with the experimental data 
available from \cite{Bachmann}.
The detector geometry in \cite{Bachmann} is same as described in Table~\ref{GEMdesign}.
The gas mixture was  $\mathrm{Ar}/\mathrm{CO_2}$ (70/30).
In Fig.~\ref{Transparency-Comparison}, the collection efficiency has 
been plotted with ${\mathrm{E_{Drift}}}$, whereas 
${\mathrm{V_{GEM}}}$ and ${\mathrm{E_{Induction}}}$ have been 
fixed to $300~\mathrm{V}$ and $2~\mathrm{kV/cm}$, respectively.
At lower drift field, for example at 
${\mathrm{E_{Drift}}}~=~0.5~\mathrm{kV/cm}$, the simulation result agrees
to experimental data within $\sim2\%$, whereas at 
${\mathrm{E_{Drift}}}~=~4~\mathrm{kV/cm}$, the agreement is within
$\sim21\%$.
The manufacturing tolerances and defects in the GEM foil, 
the uncertainties in applied voltage, the possible impurities 
in the gas mixture may be the main reasons behind the discrepancy 
between the experimental and numerical estimates.

\begin{figure}[hbt]
\centering
\subfigure[]
{\label{GEM-Eff-Drift}\includegraphics[scale=0.32]{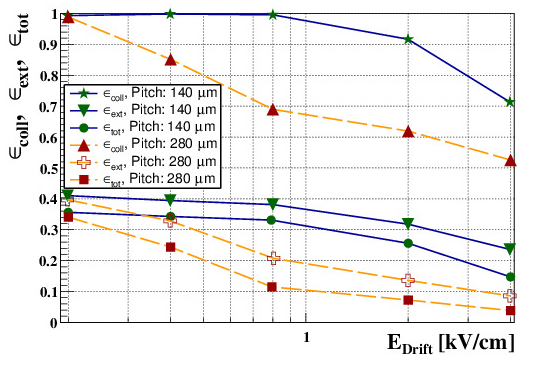}}
\subfigure[]
{\label{GEM-Eff-Ind}\includegraphics[scale=0.32]{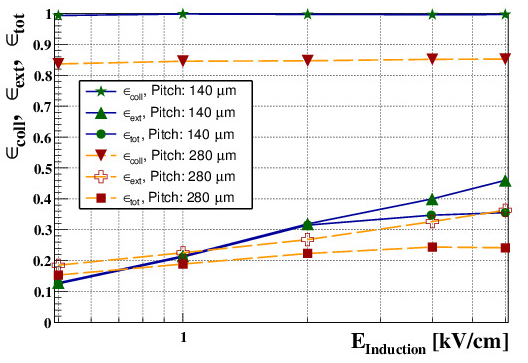}}
\subfigure[]
{\label{GEM-Eff-Vgem}\includegraphics[scale=0.32]{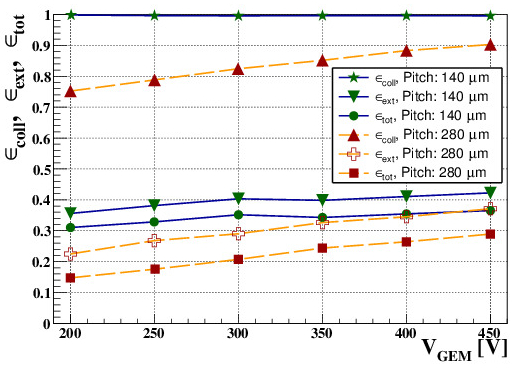}}
\subfigure[]
{\label{GEM-Eff-B}\includegraphics[scale=0.32]{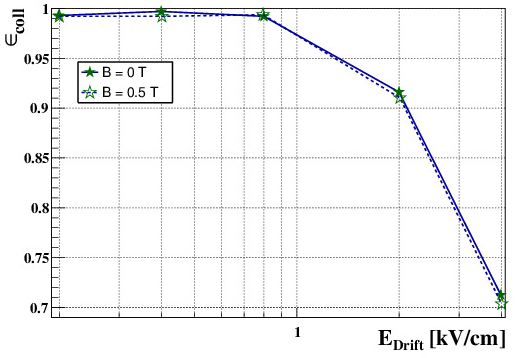}}
\caption{The variation of $\epsilon_{{coll}}$, $\epsilon_{{ext}}$ and 
$\epsilon_{{tot}}$ with (a) ${\mathrm{E_{Drift}}}$ 
(${\mathrm{V_{GEM}}} = 350~\mathrm{V}$, 
${\mathrm{E_{Induction}}} = 4~\mathrm{kV/cm}$), (b) ${\mathrm{E_{Induction}}}$ 
(${\mathrm{V_{GEM}}} = 350~\mathrm{V}$, 
${\mathrm{E_{Drift}}} = 0.4~\mathrm{kV/cm}$), 
(c) ${\mathrm{V_{GEM}}}$ (${\mathrm{E_{Drift}}} = 0.4~\mathrm{kV/cm}$, 
${\mathrm{E_{Induction}}} = 4~\mathrm{kV/cm}$). 
The effect of a magnetic field of $0.5~\mathrm{T}$ is shown in (d).}
\label{GEM-Transmission}
\end{figure}

For the configuration, discussed in this paper, the variations 
of $\epsilon_{{coll}}$, $\epsilon_{{ext}}$ and 
$\epsilon_{{tot}}$ under different field configurations have been 
plotted in Fig.~\ref{GEM-Transmission}.
For a fixed $\mathrm{E_{GEM}}$ and $\mathrm{E_{Induction}}$, $\epsilon_{{coll}}$ and thus 
$\epsilon_{{tot}}$, decrease with the increase of the drift field, 
whereas no significant effects of drift field on $\epsilon_{{ext}}$ 
has been observed (Fig.~\ref{GEM-Eff-Drift}).
Similarly, at  a fixed $\mathrm{E_{GEM}}$ and $\mathrm{E_{Drift}}$, the increase of      
induction field, increases $\epsilon_{{ext}}$ as shown in 
Fig.~\ref{GEM-Eff-Ind}.
From Fig.~\ref{GEM-Eff-Drift} and ~\ref{GEM-Eff-Ind}, it is also 
obvious that the hole pitch has a 
strong impact on $\epsilon_{{coll}}$ and $\epsilon_{{ext}}$ and, thus, 
on $\epsilon_{{tot}}$.
For the same voltage configuration, the larger pitch gives less 
$\epsilon_{{coll}}$ and, thus, $\epsilon_{{tot}}$ in comparison 
to the standard pitch of $140~\mu\mathrm{m}$.
This can be understood as follows.
For low drift fields, the voltage difference across the GEM have the 
effect of focusing the field lines towards the holes. 
However, as $\mathrm{E_{Drift}}$ increases, some of the field 
lines are attracted to the copper surface and at some point they 
end there, leading to a loss of $\epsilon_{coll}$. 
This effect takes place at lower drift fields when there is more 
space between holes, such as the case of the $280~\mu\mathrm{m}$ pitch. 
An increase in the ratio $\mathrm{E_{Drift}}/\mathrm{E_{GEM}}$ results in 
the termination of the drift lines on the top surface of the GEM foil 
leading to a loss of $\epsilon_{{coll}}$.
The extraction of electrons from the holes of the GEM increases 
with higher $\mathrm{E_{Induction}}$.
For a given $\mathrm{E_{Induction}}$ a larger value of pitch leads to 
higher $\epsilon_{{ext}}$ because $\mathrm{E_{Induction}}$ is 
relatively more uniform in this case in comparison to the configuration 
with smaller pitch. 
The change of $\mathrm{V_{GEM}}$ on electron efficiencies has been shown in  
Fig.~\ref{GEM-Eff-Vgem}.
Since the ALICE TPC will be operated in presence of a $0.5~\mathrm{T}$ 
magnetic field, the effect of such field on electron transmission has 
been studied (Fig.~\ref{GEM-Eff-B}).
The direction of this magnetic field is along positive Z-axis.
But, no significant effect on transmission, has been observed.

\begin{table}
\caption{Field configuration of quadruple GEM detector.}\label{GEMvoltage}
\begin{center}
\begin{tabular}{|c|c|}
\hline
\hline
Drift Field & 0.4 kV/cm \\
\hline
$\mathrm{E_{GEM1}}$ & 40 kV/cm \\
\hline
Transfer Field 1 & 4 kV/cm \\
\hline
$\mathrm{E_{GEM2}}$ & 35 kV/cm \\
\hline
Transfer Field 2 & 2 kV/cm \\
\hline
$\mathrm{E_{GEM3}}$ & 37 kV/cm \\
\hline
Transfer Field 3 & 0.1 kV/cm \\
\hline
$\mathrm{E_{GEM4}}$ & 45 kV/cm \\
\hline
Induction Field & 4 kV/cm\\
\hline
\end{tabular}
\end{center}
\end{table}

\subsubsection{Quadruple GEM}

From the study of single GEM detector, it is observed that higher 
electron transmission can be obtained 
with higher GEM voltage, lower drift field and higher induction field.
GEM foils with standard pitch give better electron transmission.
For the present work, the voltage configuration for quadruple GEM
detectors are listed in Table~\ref{GEMvoltage}.
The drift field is low and the induction field is high as desired.
The $\mathrm{V_{GEM}}$ has been tuned in such a way so as to keep the overall 
gain to $\sim2000$.
Also the highest voltage on the fourth GEM foil and very low field in 
Transfer Gap 3 help to reduce the ion backflow efficiently.

\begin{figure}[hbt]
\centering
\includegraphics[scale=0.32]{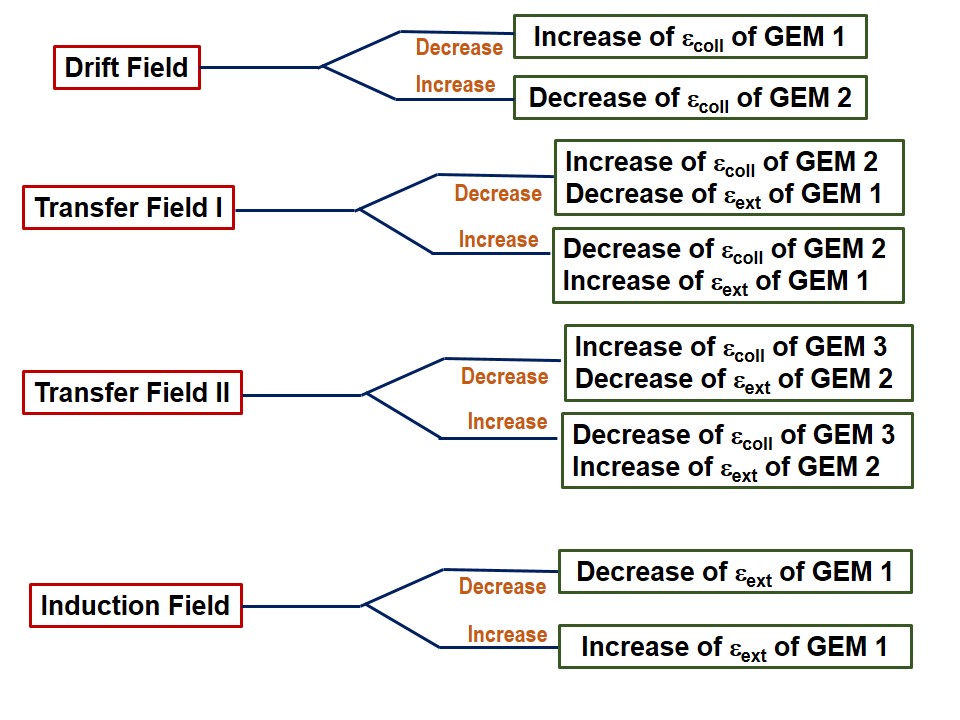}
\caption{Dependence of $\epsilon_{{coll}}$ and $\epsilon_{{ext}}$ on
different field.}
\label{GEMInformation}
\end{figure}

For multi-GEM detectors, the electron transmission can be expressed
as the multiplication of collection and extraction efficiencies of the
individual GEM foils.
Thus, for the quadruple GEM configurations, 

\begin{eqnarray}
\epsilon_{{tot}}=\epsilon_{{coll1}}\times\epsilon_{{ext1}}\times\epsilon_{{coll2}}\times\epsilon_{{ext2}}\times\epsilon_{{coll3}}\times\epsilon_{{ext3}}\times\epsilon_{{coll4}}\times\epsilon_{{ext4}}   
\end{eqnarray}

For the present voltage configuration these two efficiencies and the 
total transmission of two different quadruple GEM detectors are listed in 
Table~\ref{quadeff}.
Fig.~\ref{GEMInformation} illustrates the dependence of 
$\epsilon_{{coll}}$ and $\epsilon_{{ext}}$ on different fields.
The large field in Transfer Gap 1 and 2, which act as an induction field 
for GEM 1 and GEM 2, respectively, is sufficient for good extraction 
efficiencies from these two foils.
But, at the same time they act as a drift field for GEM 2 and GEM 3,
respectively and, thus, affect adversely the collection efficiencies of 
these two foils.
Following similar argument, the low value of Transfer Field 3 is responsible 
for the low extraction efficiency of the GEM 3 and almost $93\%$ 
collection efficiency of GEM 4.
Finally, the total transmission is affected 
significantly.
The increase of Transfer Field 2 and Transfer Field 3 and the 
decrease of the Transfer Field 
1 can affect the individual efficiencies of the GEM foil, but the 
total transmission remains unaffected.
$\epsilon_{{tot}}$ for the multi-GEM devices is also affected significantly by
the variation in geometry.

\begin{table*}
\caption{$\epsilon_{{coll}}$, $\epsilon_{{ext}}$ and $\epsilon_{{tot}}$ of
quadruple GEM detectors.}\label{quadeff}
\begin{center}
\begin{tabular}{|l|l|l|l|l|l|l|l|l|l|l|}
\hline
\hline
Geometry & B & $\epsilon_{{coll1}}$ & $\epsilon_{{ext1}}$ & $\epsilon_{{coll2}}$ & $\epsilon_{{ext2}}$ & $\epsilon_{{coll3}}$ & $\epsilon_{{ext3}}$ & $\epsilon_{{coll4}}$ & $\epsilon_{{ext4}}$ & $\epsilon_{{tot}}$ \\
& [T] & [$\%$] & [$\%$] & [$\%$] & [$\%$] & [$\%$] & [$\%$] & [$\%$] & [$\%$] & [$\%$]\\
\hline
\hline
QGemI & 0 & 99.30 & 39.56 & 6.73 & 35.43 & 15.1 & 16.02 & 91.53 & 43.98 & 0.0091 \\
\hline
QGemI & 0.5 & 99.59 & 40.02  & 6.47 & 36.16 & 14.76 & 16.08 & 90.97 & 45.49 & 0.0092 \\
\hline
QGemII & 0.5 & 89.57 & 43.09 & 7.14 & 34.59 & 12.97 & 14.26 & 97.14 & 46.10 & 0.0079 \\
\hline
\hline
\end{tabular}
\end{center}
\end{table*}

\subsection{Energy Resolution}

\begin{figure}[hbt]
\centering
\subfigure[]
{\label{Energy-Distribution}\includegraphics[scale=2.8]{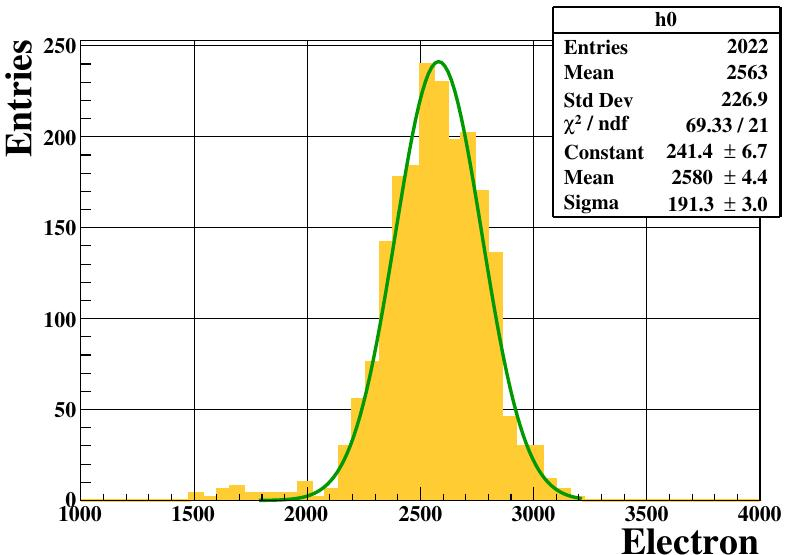}}
\subfigure[]
{\label{Energy-Drift}\includegraphics[scale=0.32]{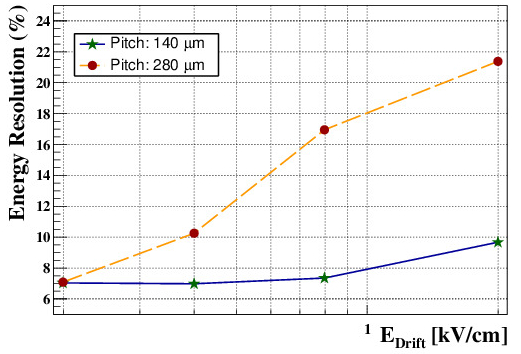}}
\subfigure[]
{\label{Energy-Ind}\includegraphics[scale=0.32]{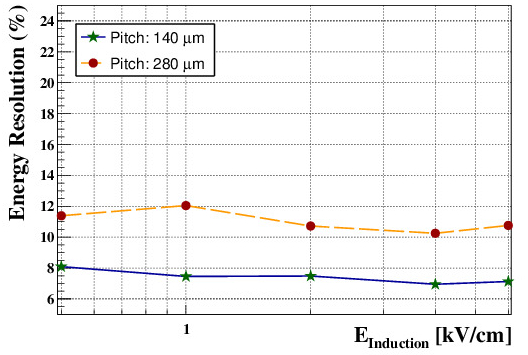}}
\subfigure[]
{\label{Energy-Vgem}\includegraphics[scale=0.32]{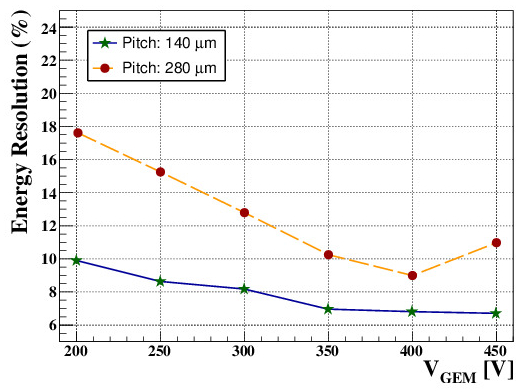}}
\caption{(a) The simulated distribution of total electron for the 5.9 keV photo-peak, the variation of energy resolution of different electrodes with (b) ${\mathrm{E_{Drift}}}$ (${\mathrm{V_{GEM}}} = 350~\mathrm{V}$, ${\mathrm{E_{Induction}}} = 4~\mathrm{kV/cm}$), (c) ${\mathrm{E_{Induction}}}$ (${\mathrm{V_{GEM}}} = 350~\mathrm{V}$, ${\mathrm{E_{Drift}}} = 0.4~\mathrm{kV/cm}$), (d) ${\mathrm{V_{GEM}}}$ (${\mathrm{E_{Drift}}} = 0.4~\mathrm{kV/cm}$, ${\mathrm{E_{Induction}}} = 4~\mathrm{kV/cm}$).}
\label{GEM-Energy}
\end{figure}

\subsubsection{Single GEM}
\label{sec:energy}

The energy resolution of a single GEM detector and its variation with 
different field configurations has been computed.
In the numerical approach, the primary ionization for $5.9~\mathrm{keV}$ 
photon track has been estimated using HEED.
These primary electrons then follow the procedures of drift and 
amplification.
Finally, the electrons have emerged from the hole and drifted towards 
the readout plane where they have been collected.
The $5.9~\mathrm{keV}$ photo-peak of the simulated charge spectrum has 
finally been fitted using a Gaussian distribution, just as it would 
have been done for an experiment.
The distribution of the total electron for the main photo-peak is 
shown in Fig.~\ref{Energy-Distribution}.
From the mean and the r.m.s of this distribution, the energy resolution 
has been estimated using
\begin{eqnarray}
{R_{energy}}~=~{{\sigma_{P}}\over{{P}}}
\end{eqnarray}

\noindent where $P$ is the peak position and the $\sigma_{{P}}$ is the r.m.s of the distribution.

The variation of energy resolution under different field configurations 
has been plotted in Fig. \ref{GEM-Energy}.
The dependence on the different fields can be explained with the help of 
the transmission plot (Fig. \ref{GEM-Transmission}).
For a fixed $\mathrm{E_{GEM}}$ and $\mathrm{E_{Induction}}$, the energy resolution is 
better at lower drift field due to the higher transmission and then it 
degrades with the increase of the drift field (Fig. \ref{Energy-Drift}).
At a fixed $\mathrm{E_{GEM}}$ and $\mathrm{E_{Drift}}$, 
the rise of induction field, increases the total transmission.
Hence, the energy resolution is better at the higher induction field 
as shown in Fig. \ref{Energy-Ind}.
On the other hand, the collection efficiency mildly depends on the GEM voltage: 
between $200~\mathrm{V}$ and $400~\mathrm{V}$ it increases only 
$\sim15\%$, while the resolution improves from $18\%$ to $9\%$. 
The increase of $\mathrm{V_{GEM}}$ by $200~\mathrm{V}$, increases the 
gain by a factor of $\sim31$, where as the standard deviation of 
single avalanche changes only by a factor of $\sim15$.
The larger amplification at high $\mathrm{V_{GEM}}$ is the main 
cause of improvement of the energy resolution (Fig. \ref{Energy-Vgem}.
But in actual experimental condition, at higher field 
(mainly at higher values of $\mathrm{E_{GEM}}$) there may be some degradation 
of the resolution due to the secondary avalanches induced by UV photons.
Besides that, in an actual experiment, the space charge and charging up of the dielectric may influence the resolution.
But, in the present calculation, we have ignored such effects for the time being.

\subsubsection{Quadruple GEM}

For numerical simulation of energy resolution, the analytical 
formula as described in the following equation, has been used.

\begin{eqnarray}\label{energyequ2}
R_{energy} = \sqrt{\frac{F}{\bar{N_P}}~+~\frac{1}{\bar{N_P}} \bigg(\frac{\sigma_{G}}{\bar{G}}\bigg)^2}
\end{eqnarray}

\noindent where $F$ is the Fano Factor, $\bar{N_P}$ is the number of primary 
ionization, $\bar{G}$ is the gain and $\sigma_{G}$ is the standard deviation 
of the single avalanche.

Numerically, the gain has been defined as the number of electrons 
reaching the anode divided by the number of primary electrons in the 
drift volume.
The numerical energy resolution, calculated using Eqn.~\ref{energyequ2}, 
for the present field configuration in case of 
QGemI was found to be $14.6\%$, whereas for QGemII, it is 
$15.8\%$. 
This is expected as the transmission in the second case is lower. 
The value agrees within $21\%$ with the reported experimental 
data of $12\%$ \cite{ALICE2}.

\subsection{Ion Backflow}

As mentioned earlier, the ions drifting back to the drift volume can 
disturb the homogeneity of the drift field and, thus, distort the 
behavior of the detector. 
The electron avalanche and the ion drift lines in case of a single 
GEM detector for a particular field configuration are shown in 
Fig.~\ref{SingleGEMIBF}. 
Most of the secondary ions are collected on the top surface of the GEM 
foil while the rest drift back to the drift volume. 
In order to prevent those ions from entering the drift volume, a proper 
optimization of the field in the drift volume, GEM hole and
induction region is necessary.

Experimentally, the backflow fraction has been defined as ratio of drift to 
anode current \cite{IBF}:

\begin{eqnarray}\label{ibf1}
\mathrm{IBF}~=~{{\mathrm{I_{Drift}}}\over{\mathrm{I_{Anode}}}}~=~{{\epsilon+1}\over{\bar{\mathrm{G}}}}
\end{eqnarray}
\noindent where $\epsilon$ is the number of back drifting ions coming from 
the amplification region, per incoming electron. 
It also includes a contribution from ions created during the ionisation 
process. 

It can be mentioned that in \cite{Bachmann}, the backflow fraction has been defined as:

\begin{eqnarray}\label{ibf2}
\mathrm{IBF}~=~{{\mathrm{I_{Drift}}}\over{{{\mathrm{I_{Drift}}}}+{\mathrm{I_{Top}}}}}
\end{eqnarray}
\noindent where $\mathrm{I_{Top}}$ is the current measured from 
the top electrode.

In the numerical approach, we noted down the number of ions collected on 
different electrodes and made an estimate of the backflow fraction using 
both equations.

\subsubsection{Single GEM}

\begin{figure}[hbt]
\centering
\subfigure[]
{\label{SingleGEMIBF}\includegraphics[scale=0.32]{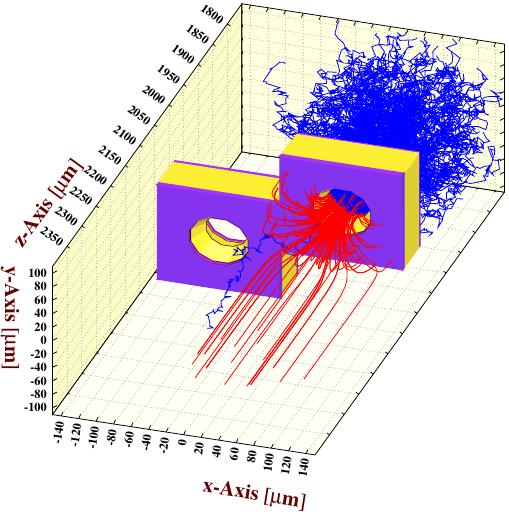}}
\subfigure[]
{\label{Ion-Comparison}\includegraphics[scale=0.32]{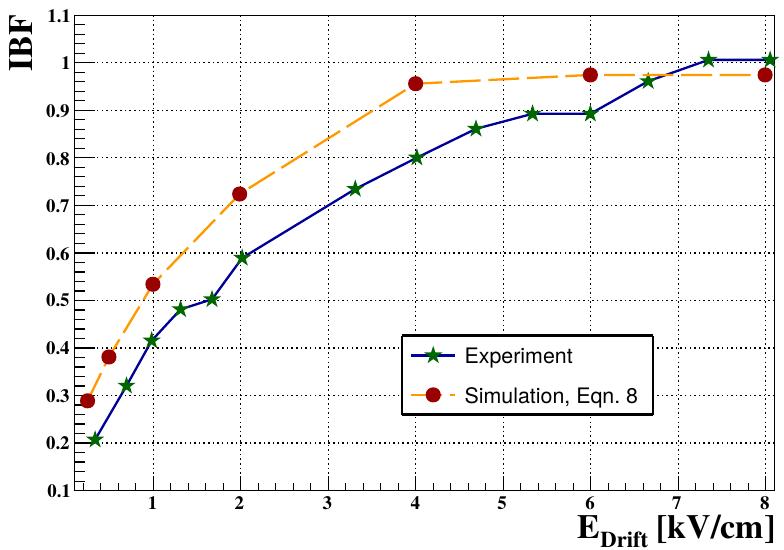}}
\caption{(a) Electron avalanche and ion drift lines for a single GEM detector. The blue lines correspond to the electron drift whereas the red ones are for ions. (b) The variation of ion backflow fraction with ${\mathrm{E_{Drift}}}$. A comparison between the experimental data from \cite{Bachmann} and simulation results using Eqn.~\ref{ibf2} is shown in (b).}
\label{IBF-GEM}
\end{figure}

We have compared our numerical estimates with the experimental data, 
from \cite{Bachmann}. 
In Fig.~\ref{Ion-Comparison}, the ion backflow fraction has been plotted 
with $\mathrm{E_{Drift}}$.
The IBF calculated using Eqn.~\ref{ibf2} agrees within $21\%$ with the 
experimental data.
Possible reasons of the discrepancy between the 
experimental and simulation results have been described 
in section ~\ref{sec:trans} and ~\ref{sec:energy}.

\begin{figure}[hbt]
\centering
\subfigure[]
{\label{Ion-Drift}\includegraphics[scale=0.32]{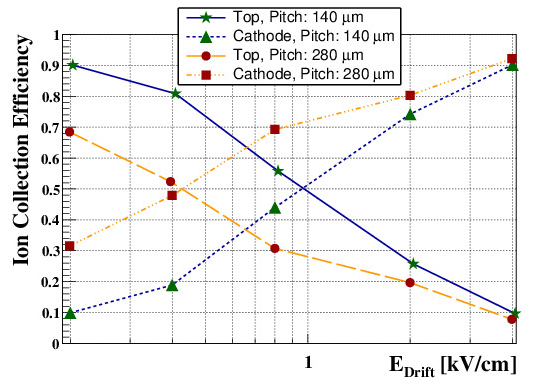}}
\subfigure[]
{\label{Ion-Ind}\includegraphics[scale=0.32]{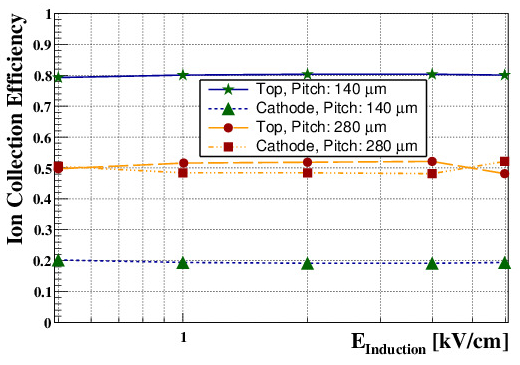}}
\subfigure[]
{\label{Ion-Vgem}\includegraphics[scale=0.32]{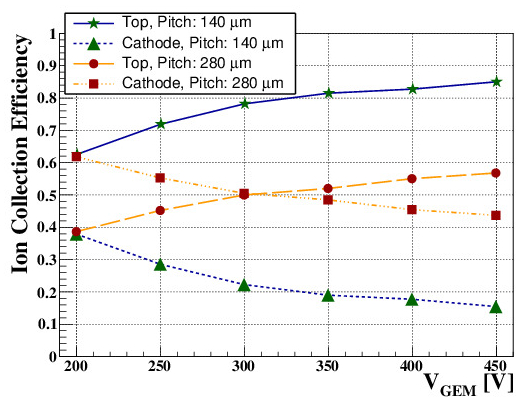}}
\subfigure[]
{\label{Ion-2Method}\includegraphics[scale=0.32]{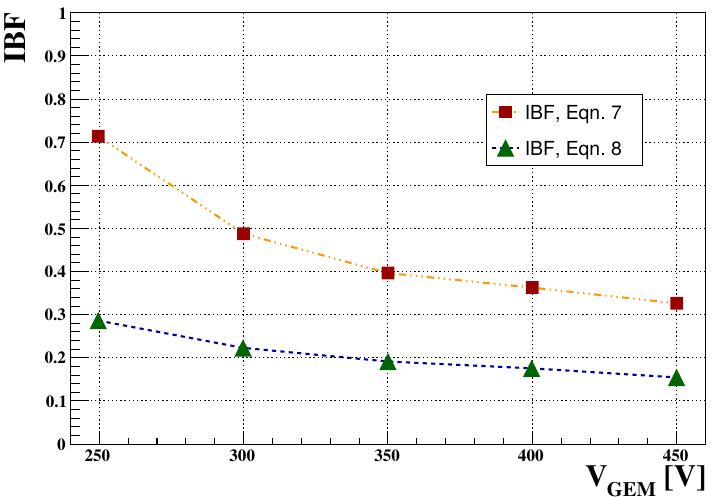}}
\caption{The variation of ion collection efficiency of different electrodes with (a) ${\mathrm{E_{Drift}}}$ (${\mathrm{V_{GEM}}} = 350~\mathrm{V}$, ${\mathrm{E_{Induction}}} = 4~\mathrm{kV/cm}$), (b) ${\mathrm{E_{Induction}}}$ (${\mathrm{V_{GEM}}} = 350~\mathrm{V}$, ${\mathrm{E_{Drift}}} = 0.4~\mathrm{kV/cm}$), (c) ${\mathrm{V_{GEM}}}$ (${\mathrm{E_{Drift}}} = 0.4~\mathrm{kV/cm}$, ${\mathrm{E_{Induction}}} = 4~\mathrm{kV/cm}$). A comparison between IBF, estimated using 
Eqn.~\ref{ibf1} and Eqn.~\ref{ibf2} is shown in (d).}
\label{Ion-Transmission}
\end{figure}

Fig.~\ref{Ion-Transmission} shows the number of ions that are collected 
on different electrode under various field configurations.
The ions that are collected on the drift plane, contribute to the 
backflow fraction.
This fraction is low when more number of ions are collected on the 
other electrode. 
For these calculations, we used  Eqn.~\ref{ibf2} in order to be 
consistent with \cite{Bachmann}.
The ion backflow of a single GEM can be reduced by decreasing $\mathrm{E_{Drift}}$ 
because less ions are extracted from the GEM holes (Fig.~\ref{Ion-Drift}).
For the same voltage configuration, for the $280~\mu\mathrm{m}$ pitch, 
due to the relatively higher drift field above the GEM foil, the ratio 
between $\mathrm{E_{Drift}}$ and $\mathrm{E_{GEM}}$ is large and, thus, the backflow 
fraction is more than that of the standard one.
No significant effect of $\mathrm{E_{Induction}}$ has been observed except at the 
higher $\mathrm{E_{Induction}}$ (Fig.~\ref{Ion-Ind}).
At higher $\mathrm{E_{GEM}}$, the ratio between $\mathrm{E_{Drift}}$ and $\mathrm{E_{GEM}}$ is 
small and, thus, a large fraction of ions is collected at the top surface of 
the GEM foil (Fig.~\ref{Ion-Vgem}).
Finally, a comparison between IBF estimated using Eqn.~\ref{ibf1} 
and Eqn.~\ref{ibf2} has been presented in Fig.~\ref{Ion-2Method}.

\subsubsection{Quadruple GEM}

A better suppression of the ion backflow is known to be achieved by using 
multiple GEM structures.
The ion collection efficiency of different electrodes under different 
configuration is listed in Table~\ref{quadion}.
The S-LP-LP-S configuration allows to block ions efficiently by
employing asymmetric transfer fields and foils with low optical transmission.
For the present voltage configuration, the gain $\sim1950$ is obtained 
with a Penning Transfer rate of $65\%$.
An increasing sequence of gas gains down the GEM stack helps reducing the ion
backflow since ions created in the inner two layers are blocked more
efficiently.
Besides that, due to the low Transfer Field 3, most of the ions created in 
the last GEM foils are collected on the top surface of this foil.
Using Eqn.~\ref{ibf2}, for the QGemI geometrical configuration, a 
backflow fraction of $2\%$ has been obtained, whereas Eqn.~\ref{ibf1}, gives 
$5.5\%$.
An increase of the Transfer Field 2 from $2~\mathrm{kV/cm}$ to 
$4~\mathrm{kV/cm}$ improves the backflow fraction by $15\%$.

\begin{table*}
\caption{Ion collection efficiency of quadruple GEM detectors.}\label{quadion}
\begin{center}
\begin{tabular}{|c|c|c|c|c|c|c|}
\hline
\hline
Geometry & B [T] & GEM1 [$\%$] & GEM2 [$\%$] & GEM3 [$\%$] & GEM4 [$\%$] & Drift [$\%$] \\
\hline
QGemI & 0 & 2.5 & 0.4 & 1.3 & 93.2 & 2.7 \\
\hline
QGemII & 0.5 & 2.3 & 0.4 & 1.3 & 93.0 & 2.8 \\
\hline
QGemII & 0.5 & 5.9 & 0.5 & 1.2 & 92.3 & 0.1 \\
\hline
\hline
\end{tabular}
\end{center}
\end{table*}

As in the case of electron transmission, geometrical variation 
of the model can affect the backflow fraction.
For QGemII, the placement of the foils are such that the collection of ions 
on the first GEM foil increases in comparison to that of the QGemI.
Therefore, the backflow fraction reduces to $0.1\%$ (using Eqn.~\ref{ibf2} and 
$0.4\%$ using Eqn.~\ref{ibf1}).

At the present voltage settings, a working point was
identified by the ALICE TPC collaboration with an ion backflow of 
about $0.7\%$ at an energy resolution of $12\%$.
This value is within the range of the values estimated by simulation 
for two different geometries, QGemI and QGemII respectively.
In the experiment, it is difficult to ensure the exact placement 
of successive GEM foils. 
This may be one of the other possible reasons of the discrepancy 
between the experimental and simulation results.
On the other hand, for the quadruple GEM, IBF 
has been numerically estimated using a single electron avalanche 
initiated in the middle of the drift region. 
In reality, experimentally measured IBF is likely to have 
contribution from ions created throughout the gas volume.
This will also lead to a difference between the experimental and the 
numerical estimates.

\section{Novel Configuration of Triple GEM Detector}

Though quadruple GEM setup is a very promising solution in terms
of backflow fraction, the electron transmission is affected adversely.
So, in parallel, a new triple GEM configuration has been studied. 

\begin{figure}[hbt]
\centering
\subfigure[]
{\label{Triple-Drift}\includegraphics[scale=0.32]{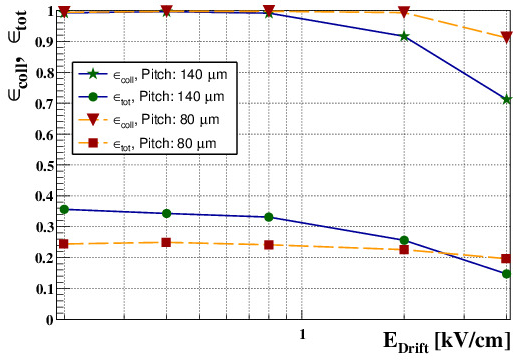}}
\subfigure[]
{\label{Triple-Induction}\includegraphics[scale=0.32]{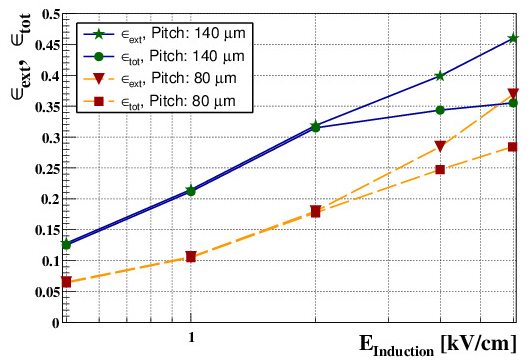}}
\subfigure[]
{\label{Triple-Ion}\includegraphics[scale=0.32]{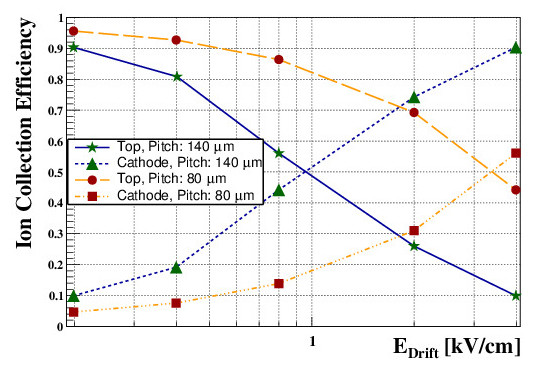}}
\caption{The variation of (a) $\epsilon_{coll}$ and $\epsilon_{tot}$ with
${\mathrm{E_{Drift}}}$, (b) $\epsilon_{ext}$ and $\epsilon_{tot}$ with
${\mathrm{E_{Induction}}}$ and (c) ion collection efficiency of different electrodes with ${\mathrm{E_{Drift}}}$ for
two different pitches having the same hole diameter.}
\label{Triple-Transmission}
\end{figure}

Earlier, in triple GEM systems using standard foils of $140~\mu\mathrm{m}$,  
ion backflow values of $4-5\%$ were observed in different gas mixtures 
\cite{IBF}.
The backflow values exceed the specifications based on the
maximum tolerable drift field distortions.
In the new configuration, a triple GEM detector,
having a configuration of LP-S-SP from top to bottom direction (here LP 
denotes the larger pitch of $280~\mu\mathrm{m}$, S stands for the standard 
pitch of $140~\mu\mathrm{m}$  and SP is the smaller pitch of 
$80~\mu\mathrm{m}$), has been proposed.
Currently, the GEM workshop at CERN produces a GEM model with 
standard active area of $100~\mathrm{cm^2}$ and a pitch of 
$90~\mu\mathrm{m}$.
The tests performed with this GEM and a detector with the geometry 
described in this section are on going and will be reported in a 
separate paper.

As seen before, the GEM collection efficiency in the same conditions 
is lower for GEMs with a larger pitch. 
Taking this into account, the aim of this setup is to apply transfer 
fields that keep a high collection efficiency for electrons in one GEM, 
while keeping a low ion collection in the holes of the previous one, 
that has a larger pitch.
The target is to get the backflow fraction less than $\sim1\%$.
Since the new configuration considers a smaller than 
standard pitch as the third GEM foil, a comparison of its characteristic 
with the standard one has first been carried out (the 
comparison between the larger pitch and the standard one has been 
already presented in the earlier section). 
This is followed by simulation of the proposed triple GEM detector 
configuration. 

\begin{figure}[hbt]
\centering
\includegraphics[scale=0.35]{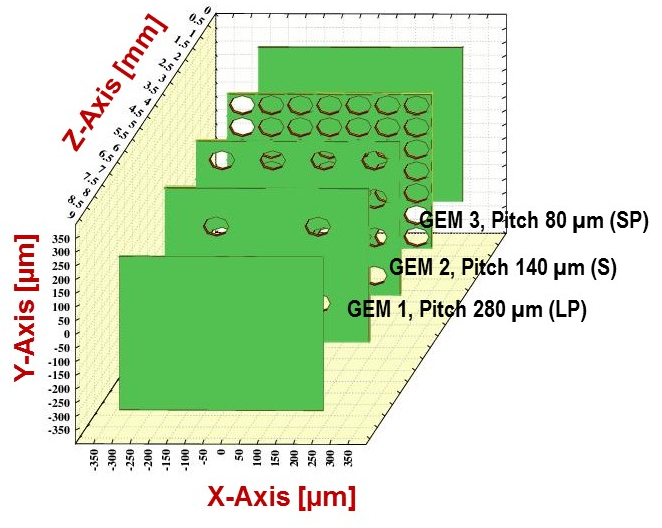}
\caption{Model of the triple GEM detector having three foils with different pitch.}
\label{Area-4}
\end{figure}

\begin{table}
\caption{Field configuration of triple GEM detector.}\label{tripleGEMvoltage}
\begin{center}
\begin{tabular}{|c|c|}
\hline
\hline
Drift Field & 0.4 kV/cm \\
\hline
$\mathrm{E_{GEM1}}$ & 52 kV/cm \\
\hline
Transfer Field 1 & 1.75 kV/cm \\
\hline
$\mathrm{E_{GEM2}}$ & 40 kV/cm \\
\hline
Transfer Field 2 & 3.6 kV/cm \\
\hline
$\mathrm{E_{GEM3}}$ & 35 kV/cm \\
\hline
Induction Field & 4 kV/cm\\
\hline
\hline
\end{tabular}
\end{center}
\end{table}

\subsection{Comparison Between Foils}

\begin{table*}
\caption{$\epsilon_{{coll}}$, $\epsilon_{{ext}}$ and $\epsilon_{{tot}}$ of triple GEM detector.}\label{trieeff}
\begin{center}
\begin{tabular}{|c|c|c|c|c|c|c|}
\hline
\hline
$\epsilon_{{coll1}}$ [$\%$] & $\epsilon_{{ext1}}$ [$\%$] & $\epsilon_{{coll2}}$ [$\%$] & $\epsilon_{{ext2}}$ [$\%$] & $\epsilon_{{coll3}}$ [$\%$] & $\epsilon_{{ext3}}$ [$\%$] & $\epsilon_{{tot}} [$\%$]$ \\
\hline
20.0 & 29.0 & 64.0 & 38.0 & 89.0 & 24.0 & 0.3 \\
\hline
\hline
\end{tabular}
\end{center}
\end{table*}

The variation of electron and ion transmission for smaller pitch are shown 
in Fig.~\ref{Triple-Transmission}.
A comparison with the standard one reveals that collection efficiency 
and ion backflow fraction are better though extraction efficiency 
is less for this smaller pitch GEM.
Due to the larger optical transparency of the smaller pitch GEM, 
the collection efficiency stays high until much higher drift fields 
because the copper area between the holes is much smaller, 
leading to less electrons lost between the holes. 
At the same time, it is necessary to apply a much higher voltage 
across the induction gap to assure the uniform field that will 
provide an extraction efficiency similar to the standard GEM. 
At the same time, this is favorable for less number of ions
to come out.

\subsection{Electron and Ion Transmission of Triple GEM Detector}

The model of the novel configuration with the triple GEM detector 
is shown in Fig.~\ref{Area-4}.
The smaller pitch acts as the last GEM foil in order to stop most of 
the ions.
The field configuration considered for the present studies is 
listed in the Table~\ref{tripleGEMvoltage}.
For the smaller pitch a relatively higher drift field is also suitable for 
a reasonably good collection efficiency.
Therefore, very low field at the transfer region is not required in this 
configuration.
For the present voltage configuration the gain is $\sim1800$ which is 
close to the ALICE requirement.
The electron transmission efficiencies and the ions collection 
efficiencies of the 
individual GEM foil are listed in Table~\ref{trieeff} and 
Table~\ref{triion}, respectively.
The total electron transmission is better than the quadruple GEM detector.
The backflow fraction is $0.2\%$ (using Eqn.~\ref{ibf2}) as desired by 
the ALICE TPC.

\begin{table}[hbt]
\caption{Ion collection efficiency of triple GEM detector.}\label{triion}
\begin{center}
\begin{tabular}{|c|c|c|c|}
\hline
\hline
GEM1 [$\%$] & GEM2 [$\%$] & GEM3 [$\%$] & Drift [$\%$] \\
\hline
8.9 & 12.8 & 77.4 & 0.2 \\
\hline
\hline
\end{tabular}
\end{center}
\end{table}

\section{Conclusion}

Time Projection Chambers (TPCs) are ideal devices
for three-dimensional tracking, momentum measurement
and identification of charged particles.
They are used in many on-going experiments, including ALICE.
In the upgraded version of the ALICE TPC, the amplification device 
will be based on the GEM detector.
The geometry proposed by the ALICE collaboration has achieved an 
excellent energy resolution with an ion backflow below $1\%$, 
while handling the proposed collision rate.
In this work, an attempt has been made to numerically model and
analyze the geometrical and electrical configuration of GEM-based 
TPCs in terms of electron and ion transmission.
Study of single GEM detectors shows that higher electron transmission, 
better energy resolution and lower backflow fraction can be obtained 
with higher GEM voltage, lower drift field and higher induction field.
GEM foils with standard pitch gives better electron transmission, as 
well as less ion backflow fraction.
No significant effect of $0.5~\mathrm{T}$ magnetic field has been 
observed on electron transmission and ion backflow fraction. 
Multi-GEM devices are found to be better in terms of lower ion backflow fraction
though the electron transmission is affected adversely.
Several studies have been performed on quadruple GEM detectors 
with various geometry and field configuration which are likely candidates 
for the TPCs in general.
Extensive comparison with the ALICE experimental data leads us to
believe that the physics processes occurring within these device are 
reasonably well understood and the tools used for carrying out the 
investigations in this work are quite mature.
Finally, numerical simulation has been performed using three GEM foils having
three different pitch but same hole diameter.
Initial calculations show that this novel configuration can be 
suitable in terms of its better electron transmission and 
less ion backflow fraction.

\section{Acknowledgment}
We would like to express our sincere gratitude to Prof. Fabio Sauli 
for his invaluable advice and encouragement to continue this work.
We thank the reviewer for the valuable comments and suggestions which helped us
to enrich the content of the paper significantly.
This work has partly been performed in the framework of the RD51 Collaboration.
We wish to acknowledge the members of the RD51 Collaboration for their 
help and suggestions.
We would also like to thank the members of the ALICE-TPC collaboration for 
their valuable suggestions.
H. Natal da Luz acknowledges FAPESP grant 2013/17405-3.
We thank our respective Institutions for providing us with the necessary 
facilities.

\end{document}